\documentclass[aps,pre,reprint]{revtex4-1}

\usepackage{amssymb}
\usepackage{amsmath}
\usepackage{subfig}
\usepackage{graphicx}
\usepackage{color}

\begin{document}
\title{Microscopic understanding of heavy-tailed return distributions in an agent-based model}
\author{Thilo A. Schmitt}
\email{thilo.schmitt@uni-due.de}
\affiliation{Faculty of Physics, University of Duisburg-Essen, Lotharstrasse 1, 47048 Duisburg, Germany}
\author{Rudi Sch\"afer}
\affiliation{Faculty of Physics, University of Duisburg-Essen, Lotharstrasse 1, 47048 Duisburg, Germany}
\author{Michael C. M\"unnix}
\affiliation{Faculty of Physics, University of Duisburg-Essen, Lotharstrasse 1, 47048 Duisburg, Germany}
\author{Thomas Guhr}
\affiliation{Faculty of Physics, University of Duisburg-Essen, Lotharstrasse 1, 47048 Duisburg, Germany}
\date{\today}

\begin{abstract}
The distribution of returns  in financial time series exhibits heavy tails.
In empirical studies, it has been found that gaps between the orders in the order book lead to large price shifts and thereby to these heavy tails. We set up an agent based model to study this issue and, in particular, how the gaps in the order book emerge.
The trading mechanism in our model is based on a \textit{double-auction order book}, which is used on nearly all stock exchanges.
In situations where the order book is densely occupied with limit orders we do not observe fat-tailed distributions. As soon as less liquidity is available, a gap structure forms which leads to return distributions with heavy tails.
We show that return distributions with heavy tails are an order-book effect if the available liquidity is constrained. This is largely independent of the specific trading strategies.
\end{abstract}

\maketitle

\section{Introduction}

A variety of stylized facts have been identified in empirical studies of financial markets~\cite{Cont2001}. A prominent example is the heavy-tailed distribution of stock price returns~\cite{Mandelbrot1963}. The precise shape of the tails has been examined in detail~\cite{Fama1965,Mandelbrot1963,Koedijk1990,Longin1996,Lux1996,Mantegna1995}.
For return intervals smaller than one day, a power-law behavior fits the data well~\cite{Plerou1999,Gopikrishnan1999}. Although the first empirical findings are roughly $50$ years old, explanations for this effect are still subject to controversial discussions~\cite{Clark1973,Arthur1996,Mandelbrot1997,Sornette1998,Lux1999,Cont2000,Challet2000,Gabaix2003,Farmer2004a}.
Following the most common reasoning, the size of orders plays a crucial role in the emergence of the non-normal distributed returns~\cite{Ying1966,Epps1976,Rogalski1978,Gabaix2003}.
In contrast, Farmer \textit{et al.} concluded from a detailed empirical investigation that the gaps between orders in the order book lead to the heavy tails~\cite{Farmer2004}. 
Stochastic modeling can only describe this and other stylized facts, but is unable to provide a deeper understanding. 

Agent-based modeling provides  means to trace back the emergence of stylized facts to the microscopic mechanisms of trading and to the traders' behavior or strategies.
In the past a variety of agent-based models were set up~\cite{Cohen1983,Kim1989,Frankel1988,Chiarella1992,Beltratti1993,Levy1994,Lux1997}. 
A prominent example is the Santa Fe Artificial Stock Market~\cite{LeBaron2002}, developed to study the emergence of trading strategies over time. In contrast there are models which are built solely to study certain aspects using small parameter sets~\cite{Bornholdt2001,Kaizoji2002,Preis2007}. We follow the latter approach; by keeping our model simple we are able to relate our parameters to empirical information.
While in many models the price formation is the result of a balance of supply and demand, the crucial mechanism in our setup is a double-auction order book. 
Our results support the view of Farmer~\textit{et al.} that the gaps in the order book are the prime reason for the heavy tails of the return distribution. 

The paper is organized as follows: In Sec.~II, we lay out our agent-based model. The emergence of heavy-tailed return distributions within our model is discussed in Sec.~III. We conclude our findings in Sec.~IV.

\section{Model}

We implement a double-auction order book and different types of traders,  each following a fixed set of rules. The traders interact via the order book by submitting market or limit orders to buy or sell stocks. The order book is the crucial mechanism where the demand for stocks meets the available supply.

The limit orders are stored in  ascending order from the cheapest buy order to the most expensive sell order. Prices are discretized due to the tick-size, i.e, only discrete price levels are present in the order book. 
Limit orders which do not trigger a trade are stored in the order book, while marketable orders are cleared immediately. 

We use discrete time steps which correspond to simulation steps. In each simulation step an arbitrary number of traders can be active. An active trader is allowed to issue one order during this time step. After the trader finishes his trading activity, the amount of time steps to his next activity is drawn from a random distribution. For these waiting times $t_\text{wt}$ we choose an exponential distribution
\begin{align}
	p(t_\text{wt}) & = \frac{ 1 }{ \mu_\text{wt} } \exp ( - t_\text{wt} / \mu_\text{wt} ) \quad \text{with} \quad \mu_\text{wt} =  c \, N \ , 
	\label{eq:wt}
\end{align}
where $N$ is the number of traders. With the scaling parameter $c$, we calibrate the number of active traders during one time step to match empirical trade frequencies. We set $c$ to achieve roughly $5.4$ trades per minute. This corresponds to the average trade frequency if we look at the top 75\% of stocks in the S\&P 500 during the year 2007.

If more than one trader is active during a single time step,  their orders are randomized to prevent serial correlations. Traders are free to choose a lifetime for their orders in accordance with their strategy. At the end of each time step orders that have expired are removed from the order book. 
By default the lifetime of an order is drawn from an exponential distribution
\begin{align}
	p(t_\text{lt}) & = \frac{ 1 }{ \mu_\text{lt} } \exp (- t_\text{lt} / \mu_\text{lt} ) \quad , 
\end{align}
which results in an average order lifetime of $\mu_\text{lt}$. In case of very small order lifetimes, the amount of limit orders in the order book becomes so small that the order book plays a minor role. In such a case the price formation is mainly driven by the specific behavior of the trader, in particular by the distribution used to determine the order price. In the following we only consider order lifetimes that are sufficiently large to avoid this problem, i.e., $\mu_\text{lt}>40$ time steps.

The most basic trader in our model is called \textsf{RandomTrader}. He does not follow a strategy, his sell and buy decisions are completely random. Buy and sell orders are issued with the same probability. The trader places his limit orders with a price drawn from a normal distribution centered around the corresponding best price, i.e., best ask in case of sell orders and best bid in case of buy orders. The size of an order $v$ is drawn from an exponential distribution,
\begin{align}
	p(v) & =  \frac{ 1 }{ \mu_\text{vol} } \exp ( - v / \mu_\text{vol}) \quad \text{with} \quad \mu_\text{vol}=10 \quad .
\end{align}
The volumes drawn from this distribution can be scaled by multiplying them with $\kappa$. In case of the \textsf{RandomTrader} the scaling parameter $\kappa$ is equal to one. The distribution and its parameter $\mu_\text{vol}$ have been chosen to roughly fit empirical order sizes found in the S\&P 500 index for the year 2007.

To study the effects of larger order sizes we use the \textsf{BigTrader}, which is a \textsf{RandomTrader} with an additional parameter $\kappa$ to scale the order size.

To avoid unnecessary boundary conditions we allow short-selling and give an unlimited credit line to each trader. While short-selling is allowed in reality we assume an unlimited credit line for the following reasons.  The individual \textsf{RandomTraders} are indistinguishable from each other due to their entirely random actions. We do not aim at a one-to-one correspondence between our simulated and real traders and it is not our goal that traders evaluate and optimize their strategy. For an external observer all incoming orders look random and therefore the actual strategies of the market participants do not play any role. The power of this approach is discussed in \cite{Farmer2005}. In this sense the random character represents the typical behavior of traders. Therefore we may assume that on average all \textsf{RandomTraders} own the same amount of money, which in our case is the average of their accumulated credit debt. As a consequence the results do not depend on the number of traders in case of the \textsf{RandomTrader}. Nevertheless a minimum of traders is required so that a wide range of waiting times, drawn from the distribution given in Eq.~(\ref{eq:wt}), is present at any time during the simulation run.

\section{Results}
We conducted the simulations with $300$ \textsf{RandomTraders}. The scale of the waiting time $c$ was adjusted so that all simulations have roughly $5.4$ trades per minute which corresponds to typical empirical time scales for  frequently traded stocks. Other values of $c$ give similar results.
\begin{figure*}[htbp]
  \begin{center}
	\subfloat{
    \includegraphics[width=0.45\textwidth]{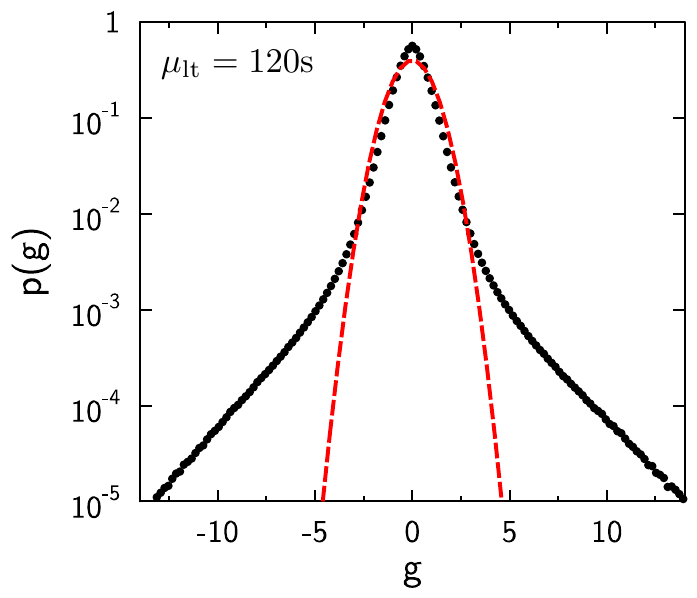}
\label{fig:results:rt_pdf:a}
}
	\subfloat{
    \includegraphics[width=0.45\textwidth]{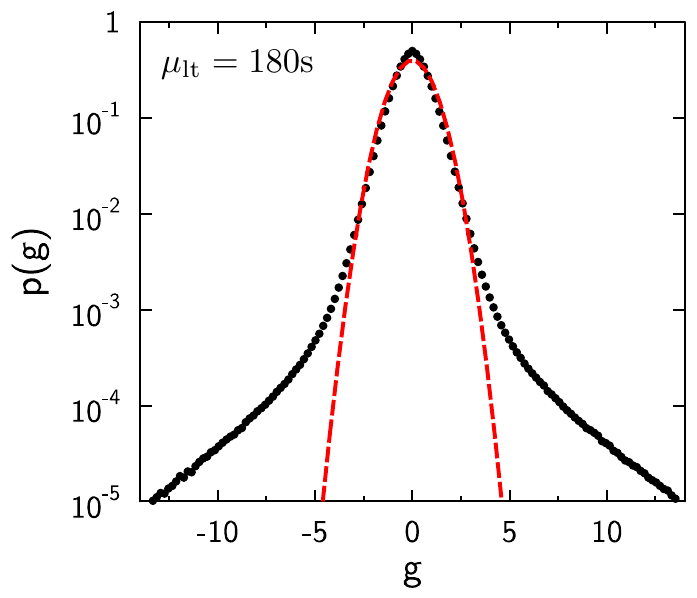}
\label{fig:results:rt_pdf:b}
}

	\subfloat{
    \includegraphics[width=0.45\textwidth]{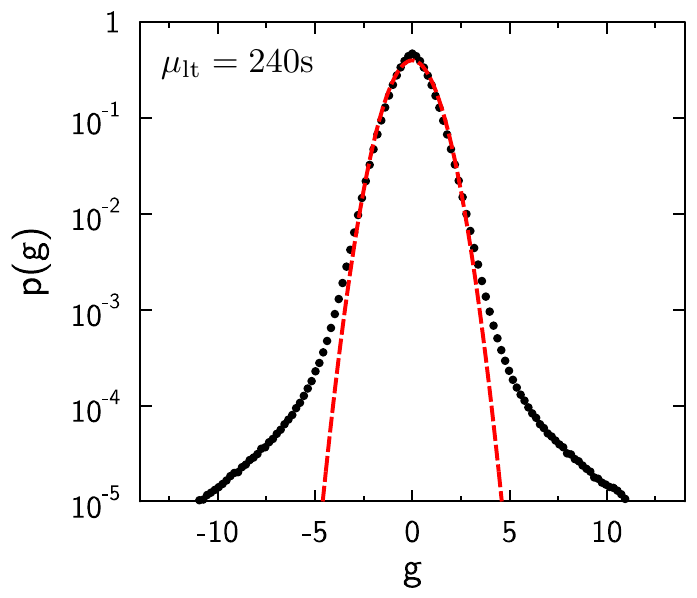}
\label{fig:results:rt_pdf:c}
}
	\subfloat{
    \includegraphics[width=0.45\textwidth]{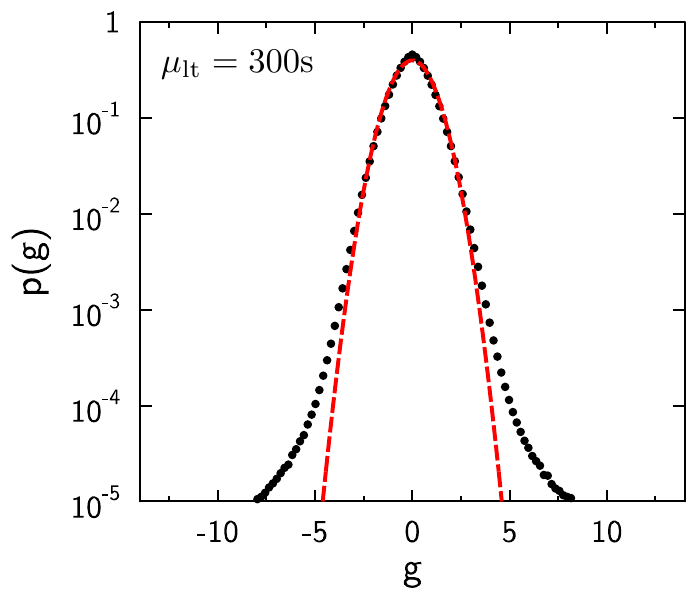}
\label{fig:results:rt_pdf:d}
}

	\subfloat{
    \includegraphics[width=0.45\textwidth]{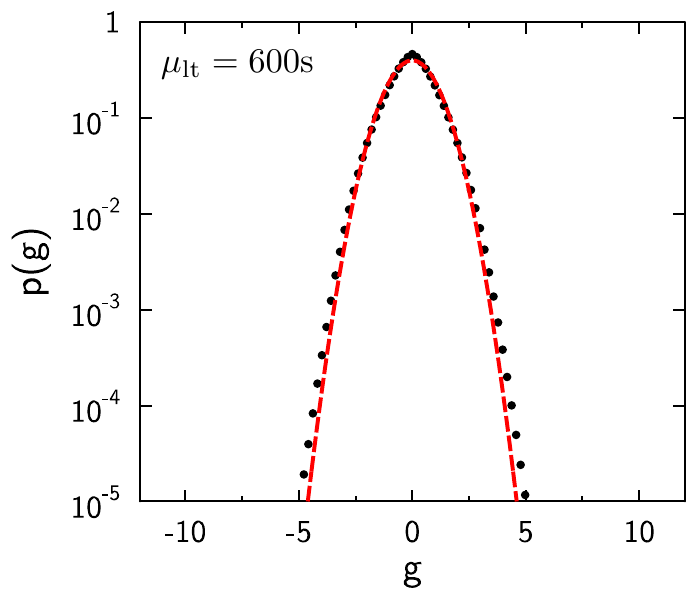}
\label{fig:results:rt_pdf:e}
}
	\subfloat{
    \includegraphics[width=0.45\textwidth]{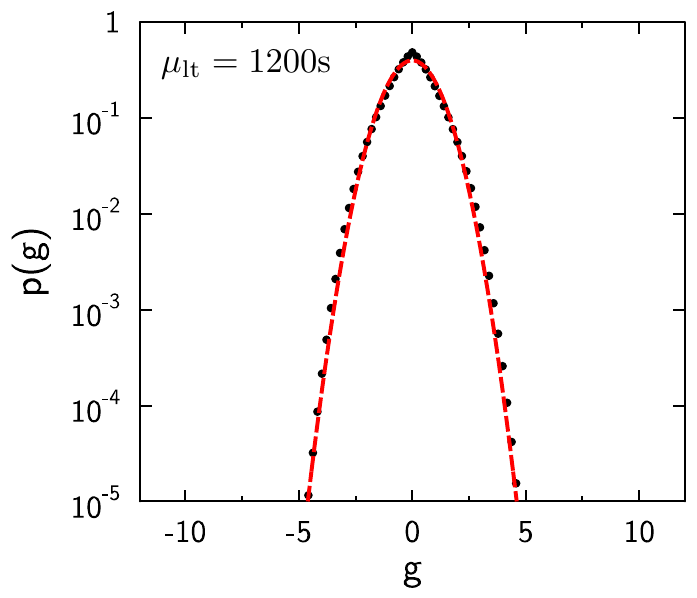}
\label{fig:results:rt_pdf:f}
}
  \end{center}
 \caption{The figures show the probability distribution of normalized one-minute returns $g$ for six simulation scenarios which differ only by their order lifetime $\mu_\text{lt}$. A normal distribution (red, dashed) is shown for comparison.}
 \label{fig:results:rt_pdf}
\end{figure*}
We use  traded prices $s(t)$ to calculate the one-minute ($\Delta t = 60$ time steps) returns
\begin{align}
r(t) = \frac{ s(t+\Delta t ) - s(t) }{s(t) }
\end{align}
for each trading day. For these returns we calculate the standard deviation
\begin{align}
	\sigma = \sqrt{ \langle r(t)^2 \rangle_T - \langle r(t) \rangle_T^2 }
\end{align}
and the mean $\mu=\langle r(t) \rangle_T $ for each simulation with a length of $T=5\cdot 10^5$ time steps. Next, we calculate the normalized returns
\begin{align}
g(t) = \frac{ r(t) - \mu }{ \sigma } \quad ,
\end{align}
which are independent of the simulation length $T$ for large $T$. It is also worth noting that the simulations are stable for the simulation length used here, i.e. the price fluctuates around the starting price for a time much greater than $T$.

In Fig.~\ref{fig:results:rt_pdf} the probability density functions for normalized returns $g$ of six simulation scenarios are shown for different values of the order lifetime $\mu_\text{lt}$.  The probability distributions are generated from $5000$ simulations with different random seeds. 
For small order lifetimes (around $40 < \mu_\text{lt} < 400$ time steps) we observe a fat-tailed return distribution, while for larger order lifetimes the return distribution approaches a normal distribution.
\begin{figure*}[htbp]
  \begin{center}
    	\includegraphics[width=0.9\textwidth]{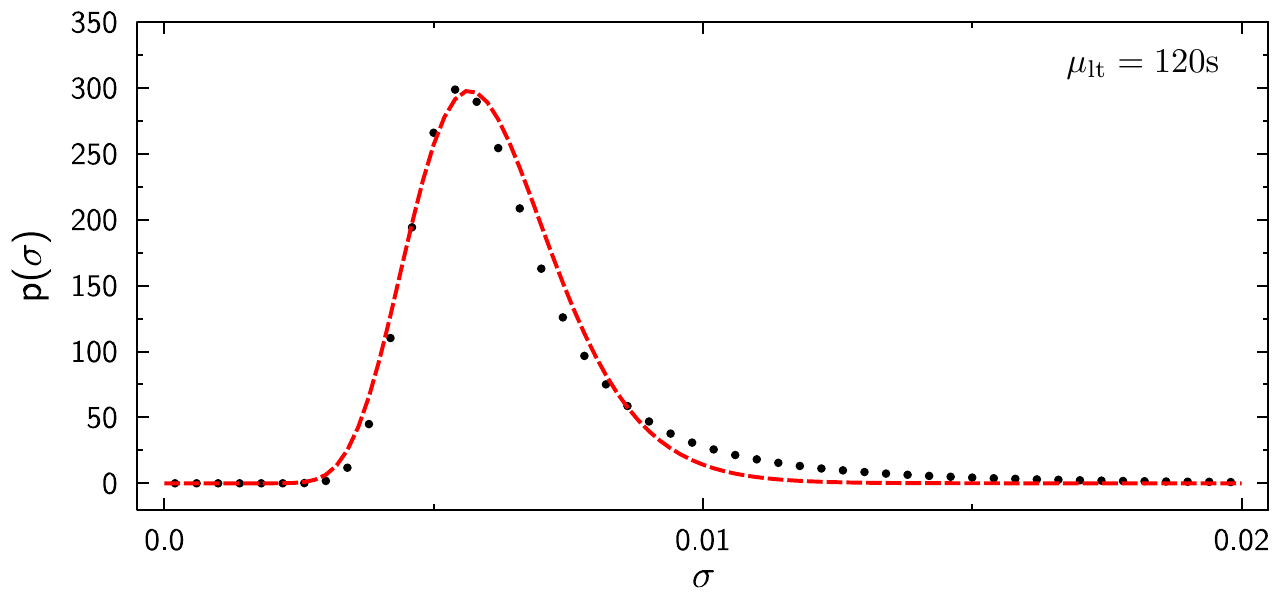}
  \end{center}
 \caption{The plot shows the probability distribution of the volatilities $p(\sigma)$. The volatilities were calculated for a moving time window of $T=1000$ time steps using one-minute returns from simulations with an order lifetime of $\mu_\text{lt}=120$. A log-normal distribution is given for comparison.}
 \label{fig:results:vola120}
\end{figure*}
The distribution of volatilities $p(\sigma)$ for an order lifetime of $\mu_\text{lt}=120$ time steps calculated for a moving window of $T=1000$ time steps is shown in Fig.~\ref{fig:results:vola120}. Previous studies showed that  empirical found volatilities are described quite well by a log-normal distribution \cite{Micciche2002,Liu1999} which is shown for comparison. Clearly, due to the simplicity of our model, we cannot expect a full agreement with the empirical distribution. It is rather an encouraging corroboration of our approach that we  get a qualitatively similar distribution.

\begin{figure*}[htbp]
  \begin{center}
	\subfloat{
    \includegraphics[width=0.9\textwidth]{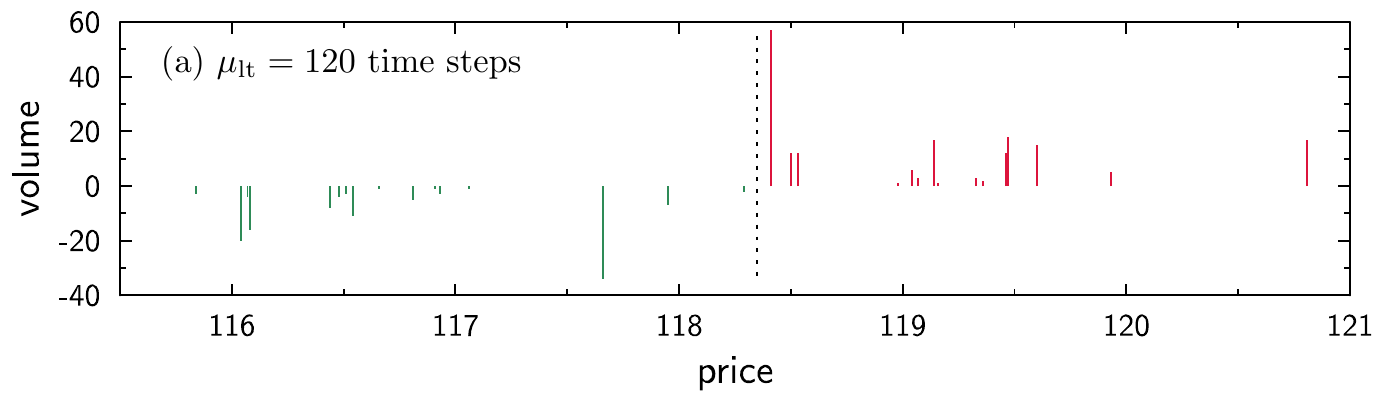}
\label{fig:results:rt_occupied:a}
}

	\subfloat{
    \includegraphics[width=0.9\textwidth]{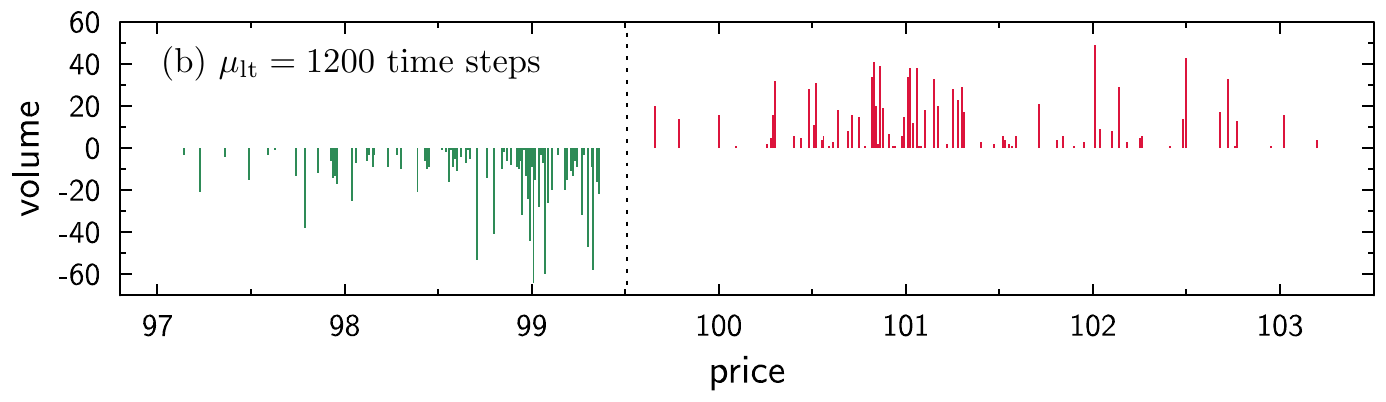}
\label{fig:results:rt_occupied:b}
}
  \end{center}
 \caption{Two typical snap shots of the order book. (a) Fewer discrete price levels are occupied and large gaps exist between the occupied levels. We assigned a negative sign to the volume of buy orders to better distinguish them from the sell orders. The midpoint is shown as a dotted line. (b) The order book is much more dense and the distance between occupied levels is much smaller compared to (a).}
 \label{fig:results:rt_occupied}
\end{figure*}

\begin{figure*}[htbp]
  \begin{center}
    	\includegraphics[width=0.9\textwidth]{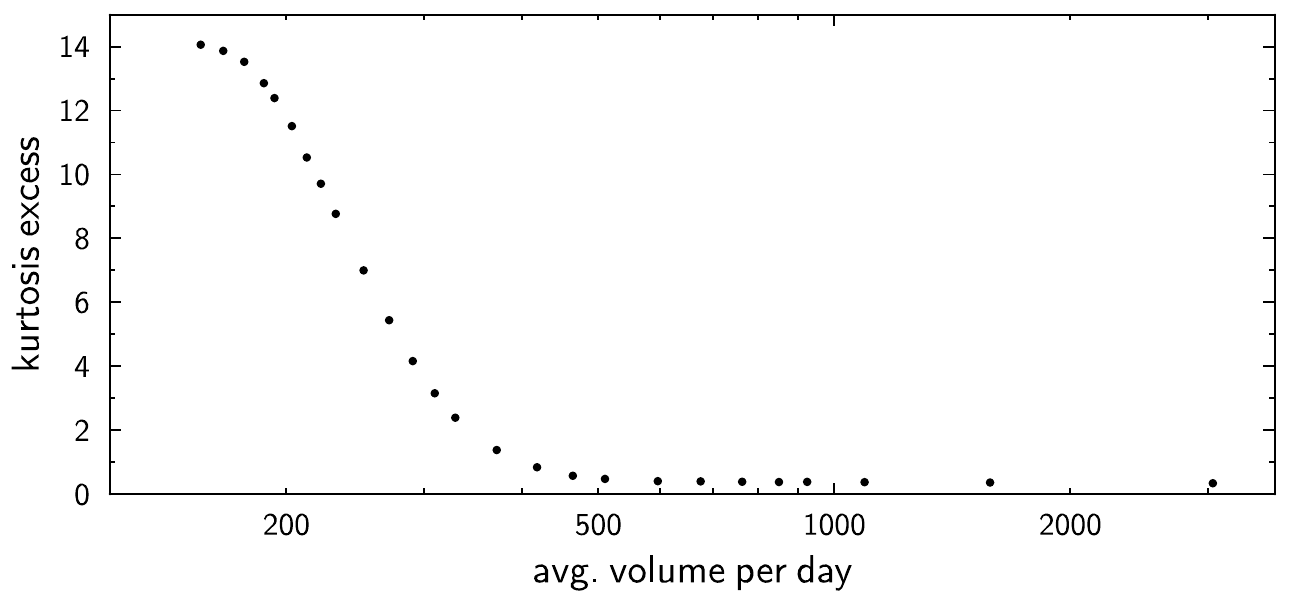}
  \end{center}
 \caption{The plot shows the excess kurtosis versus the average volume per day stored in the order book for different order lifetimes. Note that the abscissa has a logarithmic scale. The dots from left to right correspond to an order lifetime of $\mu_\text{lt}=120$, $130$, $140$, $150$, $160$, $170$, $180$, $190$, $200$, $220$, $240$, $260$, $280$, $300$, $350$, $400$, $450$, $500$, $600$, $700$, $800$, $900$, $1000$, $1200$, $1800$ and $3600$ time steps.}
 \label{fig:results:tailvsliquidity}
\end{figure*}

In our model the lifetime of an order is a way to indirectly adjust the available volume in the order book. Large order lifetimes lead to a saturation of the order book in which nearly all discrete price levels are occupied. On the other hand, for small order lifetimes less price levels are occupied. This leads to larger gaps between occupied levels, which is shown in Fig.~\ref{fig:results:rt_occupied}. We use the kurtosis excess 
\begin{align}
\gamma_2 = \frac{ \mu_4 }{ \sigma^4 } - 3 \quad ,
\end{align}
where $\mu_4$ is the fourth central moment, as a measure how much the tail of a distribution deviates from a normal distribution. For mesokurtic distributions, e.g., the normal distribution, the kurtosis excess is zero. Leptokurtic distributions whose kurtosis excess is greater than zero have more pronounced tails compared to a normal distribution.
In Fig.~\ref{fig:results:tailvsliquidity} the kurtosis excess of the return distribution is plotted versus the average volume per day in the order book. The dots from left to right correspond to different waiting times, starting with $\mu_\text{lt}=120$ and ending with $\mu_\text{lt}=3600$ time steps. We clearly see a decline of the kurtosis excess from $14$ towards zero when the volume in the order book increases. The decrease of the kurtosis excess is in line with the distributions shown in Fig.~\ref{fig:results:rt_pdf}, where the fat-tails vanish towards larger order lifetimes $\mu_\text{lt}$. We therefore note that in situations of high liquidity the fat-tails disappear.

\begin{figure*}[htbp]
  \begin{center}
	\subfloat[lifetime $\mu_\text{lt}=120$ time steps]{
    \includegraphics[width=0.46\textwidth]{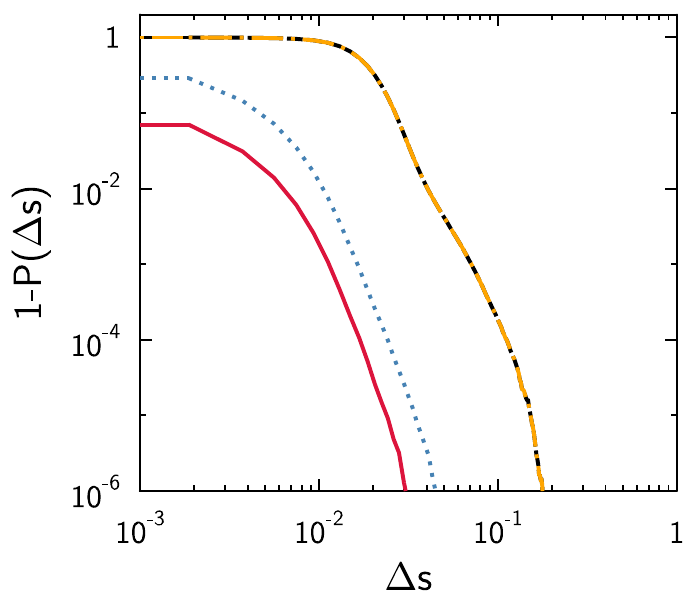}
\label{fig:results:vmi:a}
}
	\subfloat[lifetime $\mu_\text{lt}=1200$ time steps]{
    \includegraphics[width=0.46\textwidth]{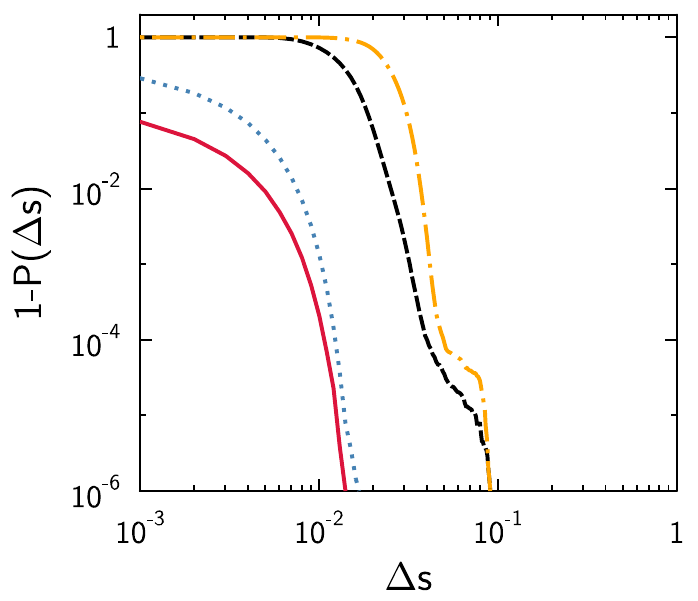}
\label{fig:results:vmi:b}
}
  \end{center}
 \caption{Complementary cumulative distribution $1-P(\Delta s)$ of the price shifts $\Delta s(t)$ for different volumes which correspond to the $0.1$ (red, solid) $0.5$ (blue, dotted) $0.9$ (black, dashed) and $0.99$ (orange, dashed-dotted) quantile of the traded volume in a double logarithmic plot. In (a) the functions for the $0.9$ and $0.99$ quantile coincide.
 }
 \label{fig:results:vmi}
\end{figure*}

To further quantify the effect of gaps within our model we study the virtual market impact function introduced in Ref.~\cite{Farmer2004}. It describes how the price would change given a hypothetical market order of volume $v$. Thus it is a test of the volume dependence of the price shift distribution. We define a supply and demand function
\begin{equation}
	\begin{split}
	S(l, t) & = \sum_{i=a(t)}^{l} V(i,t)\\
	D(l, t) & = \sum_{i=b(t)}^{l} V(i,t) \quad ,
	\end{split}
	\label{eq:theory:supplydemand}
\end{equation}
where $V(i,t)$ is the available volume at the discrete price level $i$. The supply and demand functions represent the total volume stored in the order book which is available to sell or buy up to price $l$ starting from the best ask $a(t)$ or best buy $b(t)$.
The inverse functions of $S$ and $D$, $l(S,t)$ and $l(D,t)$, are the virtual market impact functions. The inverse functions do exist because the supply and demand functions in Eq.~(\ref{eq:theory:supplydemand}) are monotonically increasing. For example, a buy market order of size $S$ will produce a price shift of $\Delta s(t)=l(S,t)-a(t)$.  After calculating the price shifts $\Delta s(t)$ for all time steps for a fixed volume $v$ we can work out the distribution and the cumulative distribution $P(\Delta s)$. By comparing distributions for different volumes $v$ we can see whether the volume $v$ changes the shape of the cumulative distribution and therefore see the influence of order size on the price shift distribution.

We now calculate the virtual market impact functions for the simulations containing only \textsf{RandomTraders} with an order lifetime of $120$ and $1200$ time steps. Figure~\ref{fig:results:rt_pdf} shows that the return distribution for $\mu_\text{lt}=120$ time steps is fat-tailed while for $\mu_\text{lt}=1200$ time steps the distribution approaches a normal distribution. We have to constrain the number of simulations to $100$ each due to the large storage requirements of order book data.
We choose the volumes $v=3,10,700,1100$ so that they represent the $0.1, 0.5, 0.9$ and $0.99$ quantiles of the traded volume given a scenario that contains $30$ \textsf{BigTraders} with $\kappa=5$.
We thus predict what would happen if we added traders who issue orders with larger sizes to the simulation. In Fig.~\ref{fig:results:vmi:a} we see the virtual market impact functions for the first case, where we observed fat-tails. If we assume power law behavior, the linear parts will be parallel to each other, i.e., they are independent of the order size $v$ and therefore are purely the result of gaps. However, for the scenario with an order lifetime of $1200$ time steps we notice that the virtual market impact functions look different for large volumes that match the $0.9$ and $0.99$ (dashed, dashed-dotted) quantile of the traded volume in Fig.~\ref{fig:results:vmi:b}. 

The different shapes in case of an order lifetime of $1200$ time steps hint at a volume dependence. If we  set up a new simulation scenario with $300$ \textsf{RandomTraders} and $30$ \textsf{BigTraders}, who trade a volume that is $\kappa > 5$ times larger than the average volume of the \textsf{RandomTrader}, we indeed observe fat-tailed return distributions. One might be tempted to conclude that large volumes can also be the cause of heavy tails, but this is not the full explanation. As the trader places his orders around the corresponding best price using a normal distribution, the probability is much higher that orders are placed near the best price than deeper in the order book. Therefore more volume is stored around the best prices, while deeper in the order book there is less volume and the gaps become larger. This is seen in Fig.~\ref{fig:results:rt_occupied:b} especially for the sell orders for prices higher than $101.5$ units. The larger orders matching the $0.9$ and $0.99$ quantile of the traded volume dig deep into the order book, reaching those lesser occupied levels and gaps. If we lower the volume multiplier $\kappa$, the orders hit less and less gaps deep in the order book, and the return distribution approaches the normal distribution.

\section{Conclusion}

In the framework of an agent-based model we identified the gap structure in the order book as the prime reason for extreme price changes.
These gaps can arise due to different reasons: older limit orders being cancelled automatically or manually, or traders placing limit orders far away from the current midpoint. 
In general, the more liquidity is provided by limit orders in the order book, the less likely are extreme price shifts -- even when market orders with large volumes are submitted.
The fat-tailed return distributions observed in empirical data reflect that, compared to the total number of shares outstanding, only very small volumes contribute to the price formation at any given time.

One mechanism which produces gaps is the finite lifetime of limit orders.
If this lifetime is comparable to the rate at which new orders are placed, a gap structure arises which leads to non-Gaussian return distributions. Further assumptions about the trader behavior are not necessary.
We demonstrated this by only considering \textsf{RandomTraders} which place their orders with a limit price drawn from a normal distribution around the best price. Still, we observe return distributions with heavy tails.

By using only traders that act randomly and do not use a strategy the prices in our model become Markovian. In this regard our traders depart from reality where prices follow a non-Markovian process \cite{Bouchaud2004}, because there is a certain amount of traders pursuing a strategy over time. While it is possible to construct such traders we prefer to keep the model simple. From this viewpoint it is remarkable and encouraging that our scenario yields the macroscopic observables in good qualitative agreement with the empirical data.
In addition, we identified situations in our model where large order volumes yield heavy-tailed return distributions. In this case, too, we can trace back the extreme price shifts to gaps that lie deep in the order book. The probability that orders are placed far away from the current best price is low and therefore not much volume exists deeper in the order book. These limit orders can only be reached by orders with very large volume.

Our results support the view of Farmer et al.~\cite{Farmer2004}.  
In low liquidity situations, i.e., with little volume in the order book, price gaps between limit orders are even relevant close to the current midpoint. Hence, even small market orders can cause large price changes. In more general terms, the less market participants are involved in the price formation process by providing limit orders to buy and sell, the more likely it becomes to find deviations from the purely diffusive price dynamics.

\begin{acknowledgments}
We thank Per Bers\'eus for pre-studies conducted in his master thesis \cite{Berseus}. Further, we express our gratitudes to
Stefan Hindel for his contributions to our agent-based model.
\end{acknowledgments}


%

\end{document}